\pdfoutput=1

\documentclass[acp]{copernicus}

\DeclareSymbolFont{bbold}{U}{bbold}{m}{n}
\DeclareSymbolFontAlphabet{\mathbbold}{bbold}

\nolinenumbers

\begin{document}

\title{Mesospheric nitric oxide model from SCIAMACHY data}

\author[1]{Stefan~Bender}
\author[2]{Miriam~Sinnhuber}
\author[1]{Patrick~J.~Espy}
\author[3]{John~P.~Burrows}

\affil[1]{Department of Physics, Norwegian University of Science and Technology, Trondheim, Norway}
\affil[2]{Institute of Meteorology and Climate Research, Karlsruhe Institute of Technology, Karlsruhe, Germany}
\affil[3]{Institute of Environmental Physics, University of Bremen, Bremen, Germany}


\runningtitle{SCIAMACHY mesosphere NO model}

\runningauthor{S.~Bender et al.}

\correspondence{Stefan~Bender (stefan.bender@ntnu.no)}

\received{23 August 2018}
\pubdiscuss{22 October 2018}
\revised{31 January 2019}
\accepted{5 February 2019}
\published{18 February 2019}


\firstpage{1}

\texlicencestatement{This work is distributed under \newline the Creative Commons Attribution 4.0 License.}
\maketitle

\begin{abstract}
  We present an empirical model for nitric oxide (\chem{NO}) in the mesosphere
  ($\approx 60$--90\,km) derived from SCIAMACHY (SCanning Imaging Absorption spectroMeter for Atmospheric CHartoghraphY) limb scan data.
  This work complements and extends the NOEM
  (Nitric Oxide Empirical Model;~\citealp{Marsh2004}) and
  SANOMA (SMR Acquired Nitric Oxide Model Atmosphere;~\citealp{Kiviranta2018})
  empirical models in the lower thermosphere.
  The regression ansatz builds on the heritage of studies
  by~\citet{Hendrickx2017} and the superposed epoch
  analysis by~\citet{Sinnhuber2016} which
  estimate \chem{NO} production from particle precipitation.

  Our model relates the daily (longitudinally) averaged \chem{NO} number densities from
  SCIAMACHY~\citep{Bender2017, Bender2017c} as a function of geomagnetic latitude
  to the solar Lyman-$\alpha$ and the geomagnetic
  AE (auroral electrojet) indices.
  We use a non-linear regression model, incorporating a finite and seasonally
  varying lifetime for the geomagnetically induced \chem{NO}.
  We estimate the parameters by finding the maximum posterior probability
  and calculate the parameter uncertainties using
  Markov chain Monte Carlo sampling.
  In addition to providing an estimate of the \chem{NO} content in the mesosphere,
  the regression coefficients indicate regions where certain processes dominate.
\end{abstract}

\introduction\label{sec:intro}  
It has been recognized in the past decades that the mesosphere
and stratosphere are coupled in various ways~\citep{Baldwin2001}.
Consequently, climate models have been evolving to extend to
increasingly higher levels in the atmosphere to improve the accuracy
of medium- and long-term predictions.
Nowadays it is not unusual that these models include the mesosphere
(40--90\,km) or the lower thermosphere (90--120\,km)~\citep{Matthes2017}.
It is therefore important to understand the processes
in the mesosphere and lower thermosphere and to find
the important drivers of chemistry and dynamics in that region.
The atmosphere above the stratosphere ($\gtrsim 40$\,km) is coupled to solar
and geomagnetic activity, also known as space weather~\citep{Sinnhuber2012a}.
Electrons and protons from the solar wind and the radiation belts
with sufficient kinetic energy
enter the atmosphere in that region.
Since as charged particles they move along the magnetic field,
this precipitation occurs primarily
at high geomagnetic latitudes.

Previously the role of \chem{NO} in the mesosphere has been
identified as an important free radical, and in this sense
a driver of the chemistry~\citep{Kockarts1980, Barth1992, Barth1995,
Roble1995, Bailey2002, Barth2009, Barth2010},
particularly
during winter when it is long-lived because of reduced photodissociation.
\chem{NO} generated in the region between 90 and 120\,km at
auroral latitudes is strongly influenced by both solar and
geomagnetic activity~\citep{Marsh2004,
Sinnhuber2011a, Sinnhuber2016, Hendrickx2015, Hendrickx2017}.
At high latitudes,
\chem{NO} is transported down to the upper stratosphere during winter,
usually down to 50\,km and occasionally down to
30\,km~\citep{Siskind2000, Randall2007, Funke2005b, Funke2014}.
At those altitudes and also in the mesosphere, \chem{NO}
participates in the ``odd oxygen catalytic cycle which depletes ozone''~\citep{Crutzen1970}.
Additional dynamical processes also result in the strong
downward transport of mesospheric air into the upper stratosphere,
such as the strong downwelling that
often occurs in the recovery phase of a sudden stratospheric warming (SSW)~\citep{Perot2014, Orsolini2017}.
This downwelling is typically associated with the formation of an elevated stratopause.

Different instruments have been measuring \chem{NO} in the mesosphere and lower
thermosphere, but at different altitudes and at different local times.
Measurements from solar occultation instruments such as Scisat-1/ACE-FTS
or AIM/SOFIE are limited in latitude and local time (sunrise and sunset).
Global observations from sun-synchronously orbiting satellites
are available from
Envisat/MIPAS below 70\,km daily and 42--172\,km every
10 days~\citep{Funke2001, Funke2005, Bermejo-Pantaleon2011}; from
Odin/SMR between 45 and 115\,km~\citep{Perot2014, Kiviranta2018};
or from Envisat/SCIAMACHY (SCanning Imaging Absorption spectroMeter for Atmospheric CHartoghraphY) between 60 and 90\,km
daily~\citep{Bender2017} and
60--160\,km every 15 days~\citep{Bender2013}.
Because the Odin and Envisat orbits are sun-synchronous,
the measurement local times are fixed to around 06:00 and 18:00 (Odin) and 10:00 and 22:00 (Envisat).
While MIPAS has both daytime and night-time measurements,
SCIAMACHY provides daytime (10:00)
data because of the measurement principle (fluorescent UV scattering;
see~\citealp{Bender2013, Bender2017}).
Unfortunately, Envisat stopped communicating in April~2012, and therefore
the data available from MIPAS and SCIAMACHY are limited to nearly
10 years from August~2002 to April~2012.
The other aforementioned instruments are still operational and
provide ongoing data as long as satellite operations continue.

Chemistry--climate models struggle to simulate the
\chem{NO} amounts and distributions in the mesosphere and lower thermosphere
(see, for example,~\citealp{Funke2017, Randall2015, Orsolini2017, Hendrickx2018}).
To remedy the situation,
some models constrain the NO content at their top layer
through observation-based parametrizations.
For example, the next generation of climate simulations
(CMIP6; see~\citealp{Matthes2017}) and other recent model
simulations~\citep{Sinnhuber2018} parametrize particle effects
as derived partly from Envisat/MIPAS \chem{NO} measurements~\citep{Funke2016}.

\subsection*{\chem{NO} in the mesosphere and lower thermosphere}\label{ssec:MLTNO}

\chem{NO} in the mesosphere and lower thermosphere is produced by
\chem{N_2} dissociation,
\begin{reaction}\label{reac:N2to2N}
\chem{N_2} + h\nu \rightarrow \chem{N(^2D)} + \chem{N(^4S)}
	\quad (\lambda < 102\,\text{nm}) \;,
\end{reaction}
followed by the reaction of the excited nitrogen atom \chem{N(^2D)}
with molecular oxygen~\citep{Solomon1982, Barth1992, Barth1995}:
\begin{reaction}\label{reac:NpO2toNOpO}
\chem{N(^2D)} + \chem{O_2} \rightarrow \chem{NO} + \chem{O} \;.
\end{reaction}
The dissociation energy of \chem{N_2}
into ground state atoms \chem{N(^4S)}
is about 9.8\,eV
($\lambda\approx127$\,nm)~\citep{Hendrie1954, Frost1956, Heays2017}.
This energy together with the excitation energy to \chem{N(^2D)}
is denoted by $h\nu$ in Reaction~\eqref{reac:N2to2N} and
can be provided by a number of sources,
most notably by auroral or photoelectrons
as well as by soft solar X-rays.

The \chem{NO} content is reduced by photodissociation,
\begin{reaction}\label{reac:NOtoNpO}
\chem{NO} + h\nu \rightarrow \chem{N} + \chem{O}
	\quad (\lambda < 191\,\text{nm}) \;,
\end{reaction}
by photoionization
\begin{reaction}\label{reac:NOtoNOppe}
\chem{NO} + h\nu \rightarrow \chem{NO}^+ + \chem{e}^-
	\quad (\lambda < 134\,\text{nm}) \;,
\end{reaction}
and by reacting with atomic nitrogen:
\begin{reaction}\label{reac:NOpNtoN2pO}
\chem{NO} + \chem{N} \rightarrow \chem{N_2} + \chem{O} \;.
\end{reaction}
\chem{N_2O} has been retrieved in the mesosphere and thermosphere
from MIPAS (see, e.g.~\citealp{Funke2008, Funke2008a}) and from
Scisat-1/ACE-FTS~\citep{Sheese2016a}.
Model--measurement studies by~\citet{Semeniuk2008} attributed
the source of this \chem{N_2O}
to being most likely the reaction
between \chem{NO_2} and \chem{N} atoms produced by
particle precipitation:
\begin{reaction}\label{reac:NpNO2toN2OpO}
\chem{N} + \chem{NO_2} \rightarrow \chem{N_2O} + \chem{O} \;.
\end{reaction}
We note that photo-excitation and photolysis at 185\,nm (vacuum UV) of
\chem{NO} or \chem{NO_2} mixtures in nitrogen, \chem{N_2}, or helium mixtures at
1\,atm leads to \chem{N_2O} formation~\citep{Maric1992}.
Both mechanisms explaining the production of \chem{N_2O} involve excited states
of \chem{NO}. Hence these pathways contribute to the loss of \chem{NO} and
potentially an additional daytime source of \chem{N_2O} in the upper atmosphere.
\chem{N_2O} acts as an intermediate reservoir at high altitudes
($\gtrapprox 90$\,km; see~\citealp{Sheese2016a}),
reacting with \chem{O}($^1$D) in two well-known channels
to \chem{N_2} and \chem{O_2} as well as to 2\chem{NO}.
However, the largest \chem{N_2O} abundances are located below 60\,km
and originate primarily from the transport of tropospheric \chem{N_2O}
into the stratosphere through the Brewer--Dobson
circulation~\citep{Funke2008a, Funke2008, Sheese2016a}
but can reach up
to 70\,km in geomagnetic storm conditions~\citep{Funke2008a, Sheese2016a}.
Both source and sink reactions indicate that
\chem{NO} behaves differently in sunlit conditions than
in dark conditions.
\chem{NO} is produced by particle precipitation at auroral latitudes,
but in dark conditions (without photolysis) it is only depleted
by reacting with atomic nitrogen (Reaction~\eqref{reac:NOpNtoN2pO}).
This asymmetry between production and depletion
in dark conditions results in different lifetimes
of \chem{NO}.

Early work to parametrize \chem{NO} in the lower thermosphere
(100--150\,km) used SNOE measurements
from March~1998 to September~2000~\citep{Marsh2004}.
With these 2.5 years of data and using empirical orthogonal functions,
the so-called NOEM (Nitric Oxide Empirical Model) estimates
\chem{NO} in the lower thermosphere as a function of the solar
f$_{10.7\,\text{cm}}$ radio flux, the solar declination angle, and
the planetary Kp index.
NOEM is still used as prior input for
\chem{NO} retrieval, for example from MIPAS~\citep{Bermejo-Pantaleon2011, Funke2012a}
and SCIAMACHY~\citep{Bender2017} spectra.
However, 2.5 years is relatively short compared to the 11-year
solar cycle, and the years 1998 to 2000 encompass a period of
elevated solar activity.
To address this, a longer time series from AIM/SOFIE was used to determine the
important drivers of \chem{NO} in the lower thermosphere (90--140\,km)
by~\citet{Hendrickx2017}.
Other recent work uses 10 years of \chem{NO} data from Odin/SMR
from 85 to 115\,km~\citep{Kiviranta2018}.
\citet{Funke2016} derived a semi-empirical model of \chem{NO_y}
in the stratosphere and mesosphere from MIPAS data.
Here we use Envisat/SCIAMACHY \chem{NO} data from the nominal limb
mode~\citep{Bender2017, Bender2017c}.
Apart from providing a similarly long time series of \chem{NO} data,
the nominal Envisat/SCIAMACHY \chem{NO} data cover the mesosphere
from 60 to 90\,km~\citep{Bender2017}, bridging the gap between
the stratosphere and lower thermosphere models.

The paper is organized as follows: we present the data used
in this work in Sect.~\ref{sec:data}.
The two model variants, linear and non-linear, are described in
Sect.~\ref{sec:regression}.
Details about the parameter and uncertainty estimation are explained
in Sect.~\ref{sec:fit}, and we present the results
in Sect.~\ref{sec:results}.
Finally we conclude our findings in Sect.~\ref{sec:conclusions}.

\section{Data}\label{sec:data}

\subsection{SCIAMACHY \chem{NO}}\label{ssec:scia.no}

We use the SCIAMACHY
nitric oxide data set version~6.2.1~\citep{Bender2017c}
retrieved
from the nominal limb scan mode ($\approx 0$--93\,km).
For a detailed instrument description, see~\citet{Burrows1995} and~\citet{Bovensmann1999},
and for details of the retrieval algorithm, see~\citet{Bender2013, Bender2017}.

The data were retrieved for the whole Envisat period
(August~2002--April~2012).
This satellite was orbiting in a sun-synchronous orbit
at around 800\,km altitude, with Equator crossing times
of 10:00 and 22:00 local time.
The \chem{NO} number densities from the SCIAMACHY nominal mode
were retrieved from the \chem{NO} gamma band emissions.
Since those emissions are fluorescent emissions excited by solar UV,
SCIAMACHY \chem{NO} data are only available for the 10:00
dayside (downleg) part of the orbit.
Furthermore, the retrieval was carried out for altitudes from 60
to 160\,km, but above approximately 90\,km, the data reflect the
scaled a priori densities from NOEM~\citep{Bender2017}.
We therefore restrict the modelling to the mesosphere below 90\,km.

We averaged the individual orbital data longitudinally on a daily basis
according to their geomagnetic latitude within 10{\degree} bins.
The geomagnetic latitude was determined according to the
eccentric dipole approximation of the
12th generation of the International Geomagnetic Reference Field
(IGRF12)~\citep{Thebault2015}.
In the vertical direction the original retrieval
grid altitudes (2\,km bins) were used.
Note that mesospheric \chem{NO} concentrations are related to geomagnetically
as well as geographically based processes, but disentangling them is beyond the
scope of the paper.
Follow-up studies can build on the method presented here and study,
for example, longitudinally resolved timeseries.

The measurement sensitivity is taken into account via the
averaging kernel diagonal elements, and days where its
binned average was below 0.002 were excluded from the timeseries.
Considering this criterion, each bin
(geomagnetic latitude and altitude)
contains about 3400~data points.

\subsection{Proxies}\label{ssec:proxies}

We use two proxies to model the \chem{NO} number densities,
one accounting for the solar irradiance variations and
one accounting for the geomagnetic activity.
Various proxies have been used or proposed to account for the
solar-irradiance-induced variations in mesospheric--thermospheric \chem{NO},
which are in particular related to the 11-year solar cycle.
The NOEM (Nitric Oxide Empirical Model;~\citealp{Marsh2004}) uses
the natural logarithm of the solar 10.7\,cm radio flux $f_{10.7}$.
More recent work on AIM/SOFIE \chem{NO}~\citep{Hendrickx2017}
uses the solar Lyman-$\alpha$ index
because some of the main production and loss processes are driven by UV photons.
Besides accounting for the long-term variation of \chem{NO} with
solar activity, the Lyman-$\alpha$ index also includes short-term
UV variations and the associated \chem{NO} production,
for example caused by solar flares.
\citet{Barth1988} have shown that the Lyman-$\alpha$ index
directly relates to the observed \chem{NO}
at low latitudes (30\degree S--30\degree N).
Thus we use it in this work as a proxy for \chem{NO}.

In the same manner as for the irradiance variations, the ``right''
geomagnetic index to model particle-induced variations of \chem{NO}
is a matter of opinion.
Kp is the oldest and most commonly used geomagnetic index;
it was, for example, used in earlier work by~\citet{Marsh2004}
for modelling \chem{NO} in the mesosphere and lower thermosphere.
Kp is derived from magnetometer stations distributed at
different latitudes and mostly in the Northern Hemisphere (NH).
However, \citet{Hendrickx2015} found that the auroral
electrojet index (AE)~\citep{Davis1966}
correlated better with SOFIE-derived \chem{NO}
concentrations~\citep{Hendrickx2015, Hendrickx2017};~\cite[see also][]{Sinnhuber2016}.
The AE index is derived from stations distributed almost evenly
within the auroral latitude band.
This distribution enables the AE index to be more closely
related to the energy input into the atmosphere at these latitudes.
Therefore, we use the auroral electrojet index (AE)
as a proxy for geomagnetically induced \chem{NO}.
To account for the 10:00 satellite sampling,
we average the hourly AE index from noon
the day before to noon on the measurement day.

It should be noted that tests using Kp (or its linear equivalent Ap) instead of AE
and using $f_{10.7}$ instead of Lyman-$\alpha$
suggested that the particular
choice of index did not lead to significantly different results.
Our choice of AE rather than Kp
and Lyman-$\alpha$ over $f_{10.7}$
is physically based and motivated
as described above.

\section{Regression model}\label{sec:regression}

We denote the number density by $x_{\text{NO}}$
as a function of
the (geomagnetic; see Sect.~\ref{ssec:scia.no}) latitude $\phi$,
the altitude $z$,
and the time (measurement day) $t$:
$x_{\text{NO}}(\phi, z, t)$.
In the following we often drop the
subscript \chem{NO} and combine the time direction
into a vector $\vec{x}$ with the $i$th entry
denoting the density at time $t_i$, such that
$x_i(\phi, z) = x(\phi, z, t_i)$.

\subsection{Linear model}\label{ssec:mod.lin}

In the (multi-)linear case, we relate the nitric oxide
number densities $x_{\text{NO}}(\phi, z, t)$ to the
two proxies, the solar Lyman-$\alpha$ index (Ly$\alpha$($t$)) and
the geomagnetic AE index (AE($t$)).
Harmonic terms with
$\omega = 2\pi\,\text{a}^{-1} = 2\pi(365.25\,\text{d})^{-1}$
account for annual and semi-annual variations.
The linear model, including a constant offset for the background density,
describes the \chem{NO} density according to Eq.~\eqref{eq:mod.lin}:
\begin{equation}\label{eq:mod.lin}
  \begin{aligned}
	x_{\text{NO}}(\phi, z, t) &= a(\phi, z)
		+ b(\phi, z)\cdot\text{Ly}\alpha(t)
		+ c(\phi, z)\cdot\text{AE}(t) \\
	& \quad {} + \sum_{n=1}^{2} \left[d_n(\phi, z)\cos(n\omega t) \right. \\
	& \qquad\qquad \left. + e_n(\phi, z)\sin(n\omega t)\right]
	\;.
  \end{aligned}
\end{equation}
The linear model can be written in matrix form for
the $n$ measurement times $t_1, \ldots, t_n$ as Eq.~\eqref{eq:mod.linmat},
with the parameter vector $\vec{\beta}$ given by
$\vec{\beta}_{\text{lin}}  = (a, b, c, d_1, e_1, d_2, e_2)^{\top} \in \mathbb{R}^7$
and the model matrix $\mathbf{X} \in \mathbb{R}^{n\times 7}$.
\begin{equation}\label{eq:mod.linmat}
  \begin{aligned}
	\vec{x}_{\text{NO}}(\phi, z)
	&=
	\left(
	\begin{matrix}
		1 & \text{Ly}\alpha(t_1) & \text{AE}(t_1) & \cos(\omega t_1) & \sin(\omega t_1)  \\
		\vdots & & & \\
		1 & \text{Ly}\alpha(t_{n}) & \text{AE}(t_{n}) & \cos(\omega t_{n}) & \sin(\omega t_{n}) 
	\end{matrix}
	\right. \\
	&\qquad
	\left.
	\begin{matrix}
		\cos(2\omega t_1) & \sin(2\omega t_1) \\
		 & \vdots \\
		\cos(2\omega t_{n}) & \sin(2\omega t_{n})
	\end{matrix}
	\right)
	\cdot
	\left(
	\begin{matrix}
		a \\
		b \\
		c \\
		d_1 \\
		e_1 \\
		d_2 \\
		e_2
	\end{matrix}
	\right)
	\\
	&= \mathbf{X} \cdot \vec{\beta}
  \end{aligned}
\end{equation}
We determine the coefficients via least squares, minimizing the
squared differences of the modelled number densities to the measured ones.

\subsection{Non-linear model}\label{ssec:mod.nonlin}

In contrast to the linear model above,
we modify the AE index by a finite lifetime $\tau$ which
varies according to season, we denote this modified version by
$\widetilde{\text{AE}}$.
We then omit the harmonic parts in the model, and the non-linear
model is given by Eq.~\eqref{eq:mod.nlin}:
\begin{equation} \label{eq:mod.nlin}
	x_{\text{NO}}(\phi, z, t) = a(\phi, z)
		+ b(\phi, z)\cdot\text{Ly}\alpha(t)
		+ c(\phi, z)\cdot\widetilde{\text{AE}}(t)
	\;.
\end{equation}
Although this approach shifts all seasonal variations to the
AE index and thus attributes them to particle-induced effects,
we found that the residual traces of particle-unrelated seasonal
effects were minor compared to the overall improvement of the fit.
Additional harmonic terms only increase the number of free
parameters without substantially improving the fit further.

The lifetime-corrected $\widetilde{\text{AE}}$ is given
by the sum of the previous 60~days' AE values,
each multiplied by an exponential decay factor:
\begin{equation} \label{eq:mod.nlin.AEtilde}
	\widetilde{\text{AE}}(t) = \sum_{t_i = 0}^{60\,\text{d}}
	\text{AE}(t - t_i)\cdot\exp\left\{ -\frac{t_i}{\tau} \right\}
	\;.
\end{equation}
The total lifetime $\tau$ is given by a constant part $\tau_0$ plus
the non-negative fraction of a seasonally
varying part $\tau_t$:
\begin{align}
	\tau &= \tau_0 +
		\begin{cases}
			\tau_t\;, & \tau_t \geq 0 \\
			0\;, & \tau_t < 0
		\end{cases}\;, \label{eq:mod.tau} \\
	\tau_t &= d\cos(\omega t) + e\sin(\omega t) \;, \label{eq:mod.tau.t}
\end{align}
where $\tau_t$ accounts for
the different lifetime during winter and summer.
The parameter vector for this model is given by
$\vec{\beta}_{\text{nonlin}} = (a, b, c, \tau_0, d, e)^{\top} \in \mathbb{R}^6$,
and we describe how we determine these coefficients and their
uncertainties in the next section.

\section{Parameter and uncertainty estimation}\label{sec:fit}

The parameters are usually estimated by maximizing the likelihood,
or, in the case of additional prior constraints, by maximizing
the posterior probability.
In the linear case and in the case of independently
identically distributed Gaussian measurement uncertainties,
the maximum likelihood solutions
are given by the usual
linear least squares solutions.
Estimating the parameters in the non-linear case is more involved.
Various methods exist, for example conjugate gradient,
random (Monte Carlo) sampling or exhaustive search methods.
The assessment and selection of the method to estimate the
parameters in the non-linear case are given below.

\subsection{Maximum posterior probability}\label{ssec:maxpost}

Because of the complicated structure of the model function
in Eq.~\eqref{eq:mod.nlin}, in particular the lifetime parts in
Eqs.~\eqref{eq:mod.tau} and~\eqref{eq:mod.tau.t},
the usual gradient methods converge slowly, if at all.
Therefore,
we fit the parameters and assess their uncertainty ranges using Markov chain
Monte Carlo (MCMC) sampling~\citep{emcee}.
This method samples probability distributions, and we apply it to
sample the parameter space putting emphasis on parameter values
with a high posterior probability.
The posterior distribution is given in the Bayesian sense
as the product of the likelihood and the prior distribution:
\begin{equation}\label{eq:posterior}
  p(\vec{x}_\text{mod} | \vec{y}) \propto
	p(\vec{x}_\text{mod} | \vec{y}, \vec{\beta}) p(\vec{\beta}) \;.
\end{equation}
We denote the vector of the measured densities by $\vec{y}$
and the modelled densities by $\vec{x}_{\text{mod}}$ similar
to Eqs.~\eqref{eq:mod.lin} and~\eqref{eq:mod.nlin}.
To find the best parameters $\vec{\beta}$ for the model,
we maximize 
$\log p(\vec{x}_\text{mod} | \vec{y})$.

The likelihood $p(\vec{x}_\text{mod} | \vec{y}, \vec{\beta})$
is in our case given by a Gaussian distribution of the residuals,
the difference of the model to the data, given in Eq.~\eqref{eq:likelihood}:
\begin{equation}\label{eq:likelihood}
	\begin{aligned}
		p(\vec{x}_\text{mod} | \vec{y}, \vec{\beta})
		&= \mathcal{N} \left(\vec{y}, \mathbf{S}_{y}\right) \\
		&= C \exp\left\{ -\frac{1}{2}
			\left(\vec{y} - \vec{x}_\text{mod}(\vec{\beta})\right)^\top \right. \\
		& \qquad \qquad \qquad \left. \vphantom{\frac{1}{2}}
			\mathbf{S}_{y}^{-1} \left(\vec{y} - \vec{x}_\text{mod}(\vec{\beta})\right)
			\right\} \;.
	\end{aligned}
\end{equation}
Note that the normalization constant $C$ in Eq.~\eqref{eq:likelihood}
does not influence the value of the maximal
likelihood.
The covariance matrix $\mathbf{S}_{y}$ contains the squared standard errors
of the daily zonal means on the diagonal,
$\mathbf{S}_{y} = \text{diag}(\sigma_y^2)$.

The prior distribution $p(\vec{\beta})$ restricts the parameters
to lie within certain ranges, and
the bounds we used for the sampling are listed in
Table~\ref{tab:paramspace}.
Within those bounds
we assume uniform (flat) prior distributions
for the offset, the geomagnetic and solar amplitudes, and in the
linear case also for the annual and semi-annual harmonics.
We penalize large lifetimes using an
exponential distribution $p(\tau) \propto \exp\{- \tau / \sigma_\tau\}$
for each lifetime parameter, i.e.\ for $\tau_0$, $d$, and $e$ in
Eqs.~\eqref{eq:mod.tau} and~\eqref{eq:mod.tau.t}.
The scale width $\sigma_\tau$ of this exponential distribution
is fixed to 1 day.
This choice of prior distributions for the lifetime parameters
prevents sampling of the edges of the
parameter space
at places with small geomagnetic coefficients.
In those regions the lifetime may be ambiguous and less meaningful.

\begin{table}
	\caption{Parameter search space for the non-linear model and
	uncertainty estimation.}\label{tab:paramspace}
\begin{tabular}{lrrc}
	\tophline
	Parameter & Lower bound & Upper bound & Prior \\
		& & & form \\
	\middlehline
	Offset ($a$) & $-10^{10}\,$cm$^{-3}$ & $10^{10}\,$cm$^{-3}$ & flat \\
	Lyman-$\alpha$ amplitude ($b$) & $-10^{10}\,$cm$^{-3}$ & $10^{10}\,$cm$^{-3}$ & flat \\
	AE amplitude ($c$) & $0\,$cm$^{-3}$ & $10^{10}\,$cm$^{-3}$ & flat \\
	$\tau_0$ & $0\,$d & $100\,$d & exp \\
	$\tau$ cosine amplitude ($d$) & $-100\,$d & $100\,$d & exp \\
	$\tau$ sine amplitude ($e$) & $-100\,$d & $100\,$d & exp \\
	\bottomhline
\end{tabular}
\end{table}

\subsection{Correlations}\label{ssec:corr}

In the simple case, the measurement covariance matrix $\mathbf{S}_{y}$
contains the measurement uncertainties on the diagonal,
in our case the (squared) standard error of the zonal means
denoted by $\sigma_y$,
$\mathbf{S}_{y} = \text{diag}(\sigma_y^2)$.
However, the standard error of the mean
might underestimate the true uncertainties.
In addition, possible correlations may occur which
are not accounted for using a diagonal $\mathbf{S}_{y}$.

Both problems can be addressed by adding a
covariance kernel $\mathbf{K}$ to $\mathbf{S}_y$.
Various forms of covariance kernels can be
used~\citep{Rasmussen2006},
depending on the underlying process leading to the
measurement or residual uncertainties.
Since we have no prior knowledge about the true correlations,
we use a commonly chosen
kernel of the Mat\'{e}rn-3/2
type~\citep{Matern1960, MacKay2003, Rasmussen2006}.
This kernel only depends on the (time) distance between
the measurements $t_{ij} = \lvert t_i - t_j \rvert$
and has two parameters, the ``strength'' $\sigma$
and correlation length $\rho$:
\begin{equation}\label{eq:corr.mat32}
	K_{ij} = \sigma^2
		\left( 1 + \frac{\sqrt{3}\,t_{ij}}{\rho} \right)
		\exp\left\{- \frac{\sqrt{3}\,t_{ij}}{\rho} \right\}
	\;.
\end{equation}
Both parameters are estimated together with the model
parameter vector $\vec{\beta}$.
We found that using the kernel~\eqref{eq:corr.mat32}
in a covariance matrix $\mathbf{S}_y$
with the entries
\begin{equation}\label{eq:covariance}
	{S_y}_{ij} = K_{ij} + \delta_{ij} {\sigma_y}_i^2
	\;,
\end{equation}
worked best and led to stable and reliable parameter sampling.
Note that an additional ``white noise'' term
$\sigma^2 \mathbbold{1}$ could be added to
the covariance matrix to account for still
underestimated data uncertainties.
However, this additional white noise term did not
improve the convergence, nor did it influence the fitted
parameters significantly.

The approximately $3000\times3000$ covariance matrix
of the Gaussian process model for the residuals was
evaluated using the~\citet{celerite} approximation
and the provided Python code~\citep{celeritev030}.
For one-dimensional data sets, this approach is computationally
faster than the full Cholesky decomposition,
which is usually used to invert the covariance matrix $\mathbf{S}_y$.
With this approximation, we achieved sensible Monte Carlo
sampling times to facilitate evaluating all $18\times16$
latitude\,$\times$\,altitude bins on a small cluster in about 1 day.
We used the \texttt{emcee} package~\citep{emcee} for the Monte Carlo sampling,
set up to use
112~walkers and 800~samples for the initial fit of the parameters,
followed by another 800~so-called burn-in samples and 1400~production
samples.
The full code can be found at \texttt{https://github.com/st-bender/sciapy}~\citep{Bender2018a}.

\section{Results}\label{sec:results}

We demonstrate the parameter estimates using example time series
$\vec{x}_{\text{NO}}$ at 70\,km at 65\degree S, 5\degree N, and 65\degree N.
\chem{NO} shows different behaviour in these regions,
showing the most variation with respect to
the solar cycle and geomagnetic activity at high latitudes.
In contrast, at low latitudes the geomagnetic influence should be
reduced~\citep{Barth1988, Hendrickx2017, Kiviranta2018}.
We briefly only show the results for the linear model and point out
some of its shortcomings.
Thereafter we show the results from the non-linear model and
continue to use that for further analysis of the coefficients.

\subsection{Time series fits}\label{ssec:res.fit}

\begin{figure*}[t]
\centering
\includegraphics[width=5.5cm]{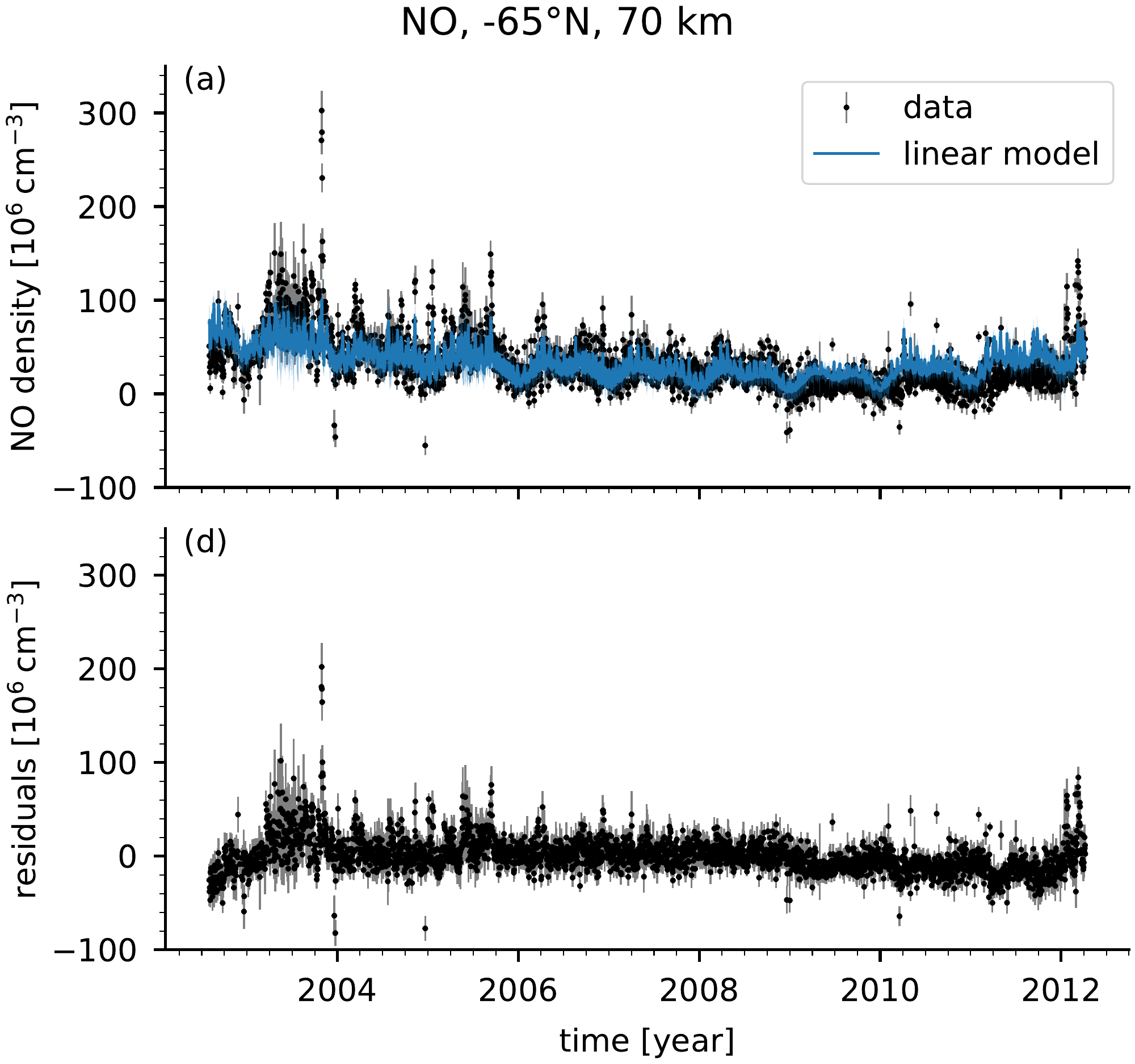}
\includegraphics[width=5.5cm]{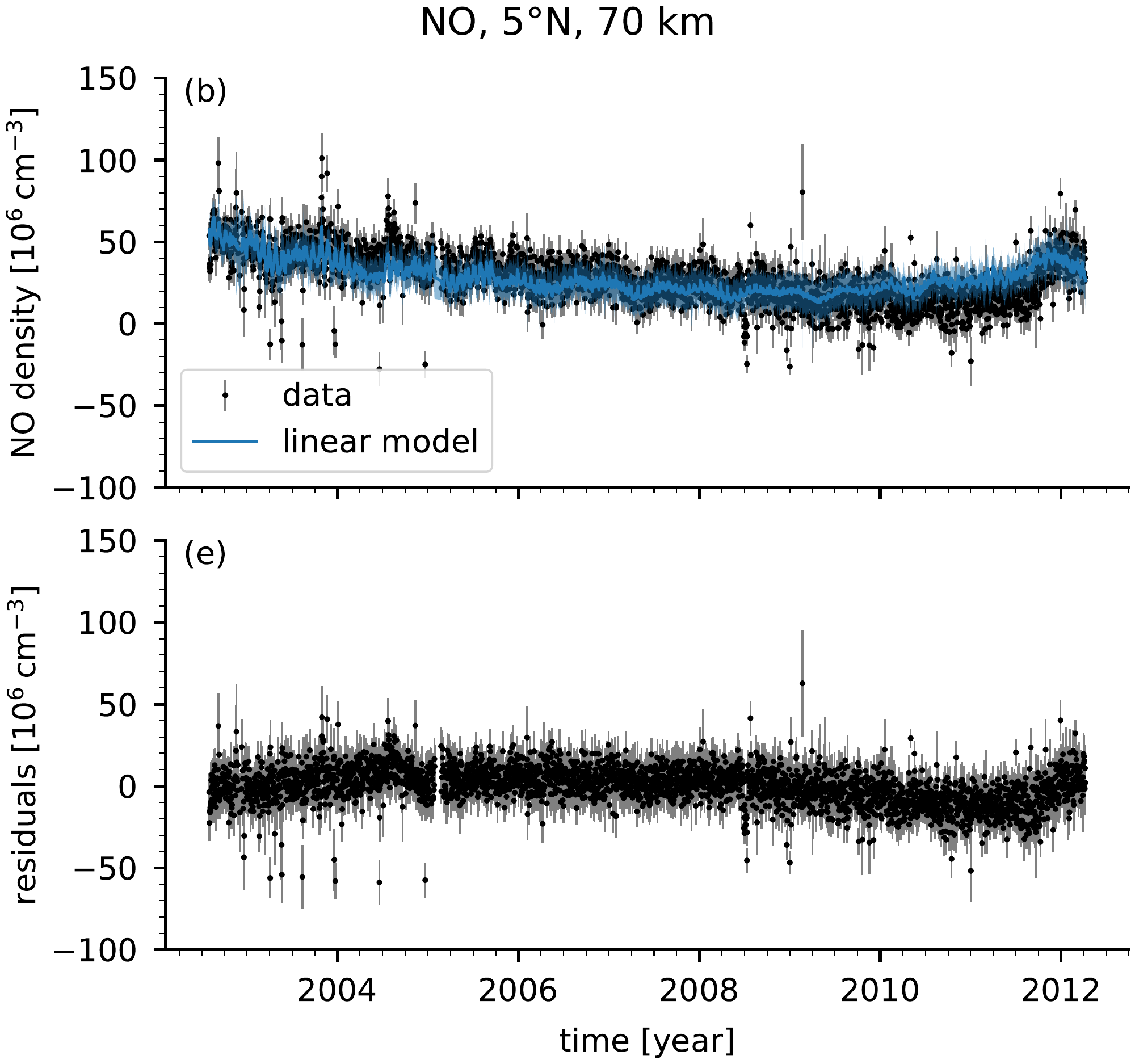}
\includegraphics[width=5.5cm]{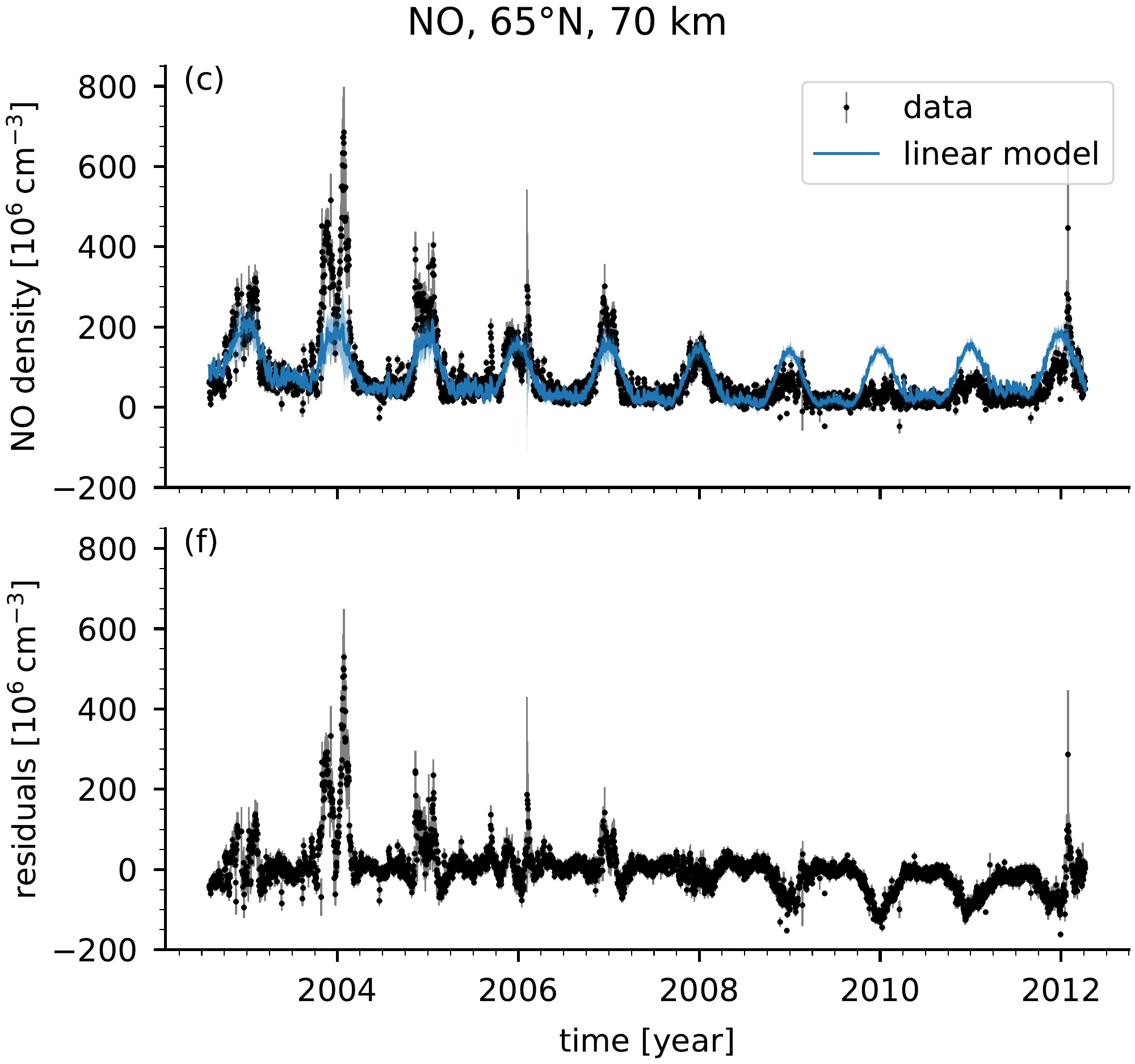}
\caption{Time series data and linear model values and residuals
	at 70\,km for 65\degree S~\textbf{(a, d)}, 5\degree N~\textbf{(b, e)}, and 65\degree N~\textbf{(c, f)}.
	Panels \textbf{(a)}--\textbf{(c)}~show the data (black dots with $2\sigma$ error bars) and the
	model values (blue line).
	Panels \textbf{(d)}--\textbf{(f)}~show the residuals as black dots with $2\sigma$ error bars.
}\label{fig:lin.fit}
\end{figure*}

The fitted densities of the linear model Eq.~\eqref{eq:mod.lin}
compared to the data are shown in the upper panels of Fig~\ref{fig:lin.fit}
for the three example latitude bins (65\degree S, 5\degree N, 65\degree N) at 70\,km.
The linear model works well at high southern and low latitudes.
At high northern latitudes and to a lesser extent at high southern latitudes,
the linear model captures the summer \chem{NO} variations well.
However, the model underestimates the high values in the polar winter
at active times (2004--2007) and overestimates the low winter values
at quiet times (2009--2011).

\begin{figure*}[t]
\centering
\includegraphics[width=5.5cm]{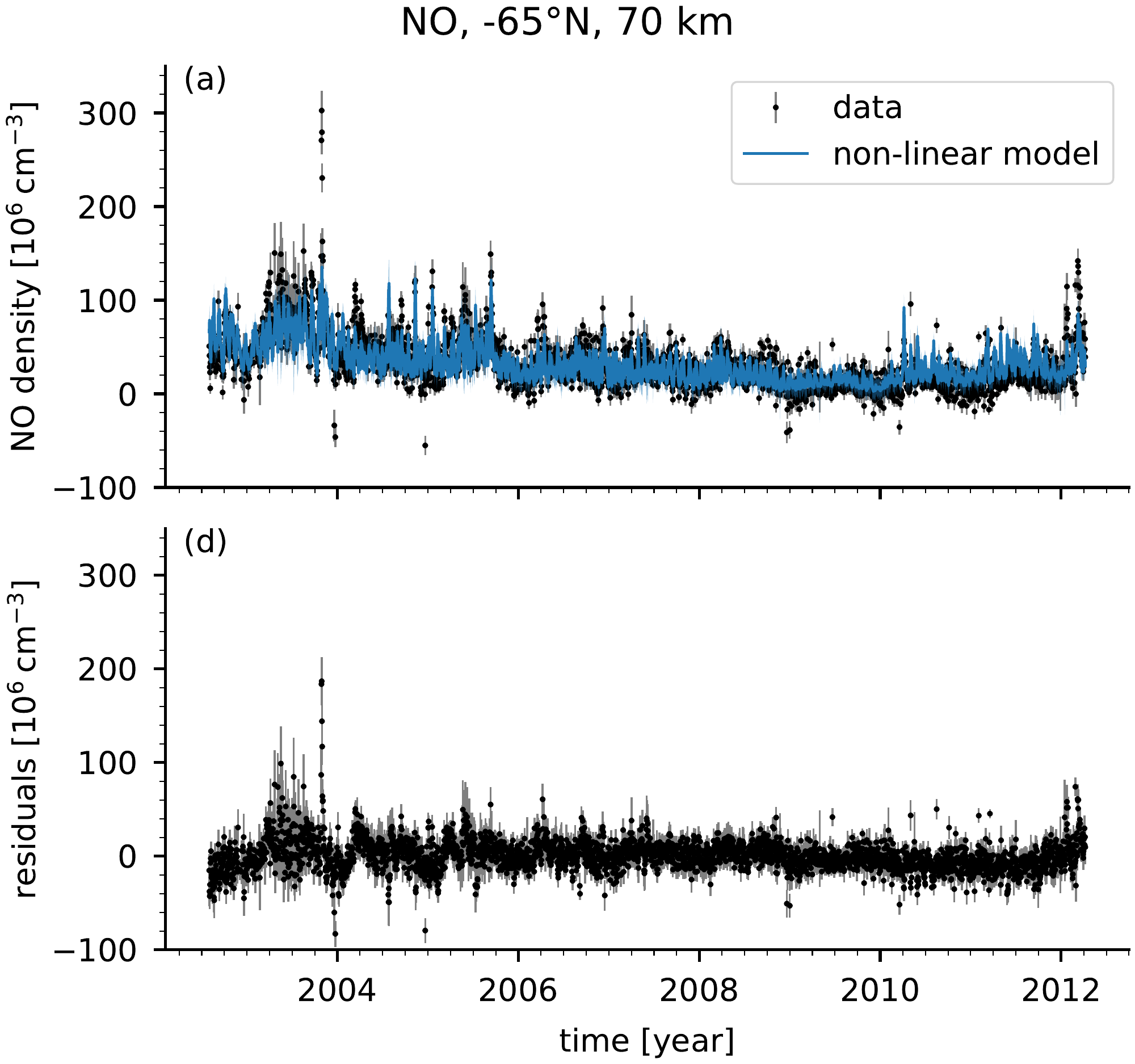}
\includegraphics[width=5.5cm]{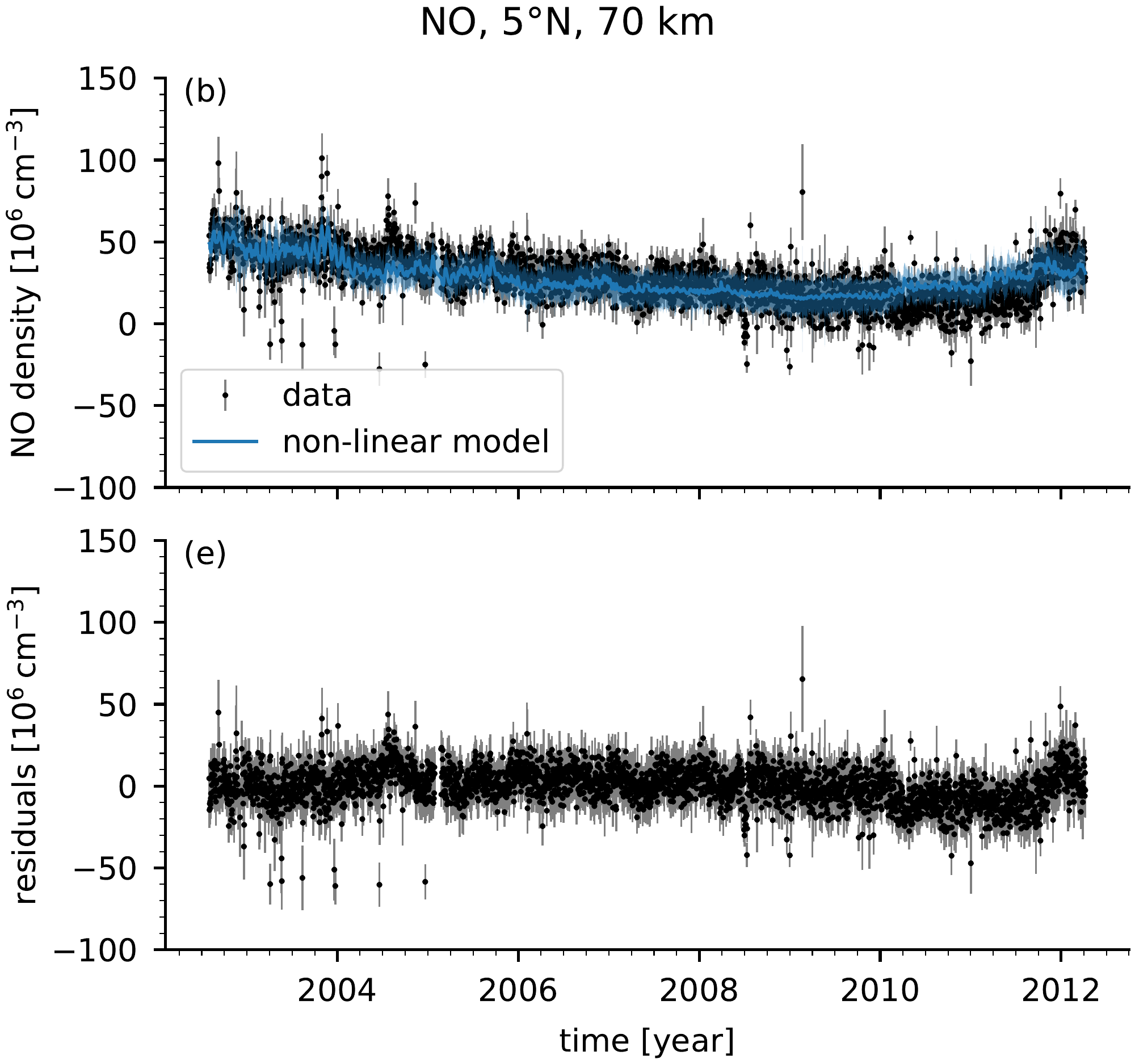}
\includegraphics[width=5.5cm]{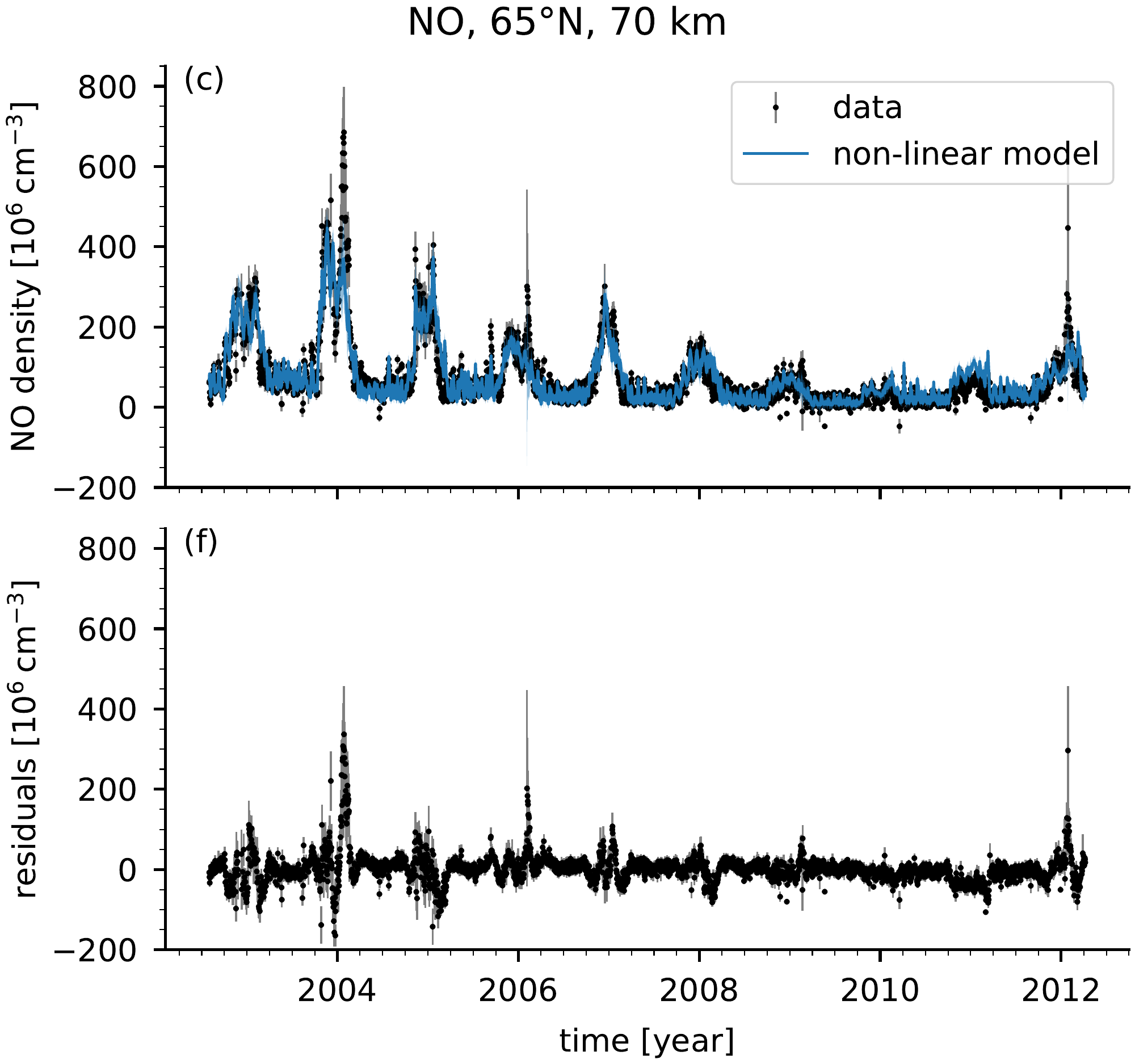}
\caption{Same as Fig.~\ref{fig:lin.fit} for the non-linear model.
}\label{fig:nonlin.fit}
\end{figure*}

For the sample timeseries (65\degree S, 5\degree N, 65\degree N at 70\,km),
the fits using the non-linear model Eq.~\eqref{eq:mod.nlin}
are shown in the upper panels of Fig~\ref{fig:nonlin.fit}.
The non-linear model better captures both the summer \chem{NO} variations as
well as the high values in the winter, especially at high northern latitudes.
However, at times of high solar activity (2003--2006) and
in particular at times of a strongly disturbed mesosphere (2004, 2006, 2012),
the residuals are still significant.
At high southern and low latitudes, the improvement over the linear model is
less evident.
At low latitudes, the \chem{NO} content is apparently mostly related
to the eleven-year solar cycle and the particle influence is
suppressed.
Since this cycle is covered by the Lyman-$\alpha$ index,
both models perform similarly, but the non-linear version
has one less parameter.
In both regions the residuals show traces of seasonal variations
that are not related to particle effects.
The linear model appears to capture these variations better than the non-linear model.
However, by objective measures including the number of model parameters%
\footnote{Past and recent research in model selection provides a number of
	choices on how to compare models objectively.
	The results are so-called information criteria which aim to provide a
	consistent way of how to compare models,
	most notably the ``Akaike information criterion'' (AIC;~\citealp{Akaike1974}),
	the ``Bayesian information criterion'' or ``Schwarz criterion'' (BIC or SIC;~\citealp{Schwarz1978}),
	the ``deviance information criterion'' (DIC;~\citealp{Spiegelhalter2002, Ando2011}),
	or the ``widely applicable information criterion'' (WAIC;~\citealp{Watanabe2010, Vehtari2016}).
	Alternatively, the ``standardized mean squared error'' (SMSE)
	or the ``mean standardized log loss'' (MSLL)~\citep[chap.~2]{Rasmussen2006}
	give an impression of the quality of regression models with respect to each other.
},
the non-linear version fits the data better in all bins (not shown here).
At high southern latitudes, the SCIAMACHY data are less
densely sampled compared to high northern latitudes
(see~\citealp{Bender2017}).
In addition to the sampling differences, geomagnetic latitudes encompass
a wider geographic range in the Southern Hemisphere (SH)
than in the Northern Hemisphere (NH),
and the AE index is derived from stations in the NH.
Both effects can lower the \chem{NO} concentrations that SCIAMACHY
observes in the SH, particularly at the winter maxima.
The lifetime variation that improves the fit in the NH
is thus less effective in the SH.

\subsection{Parameter morphologies}\label{ssec:res.morph}

\begin{figure*}[t]
\centering
\includegraphics[width=16.6cm]{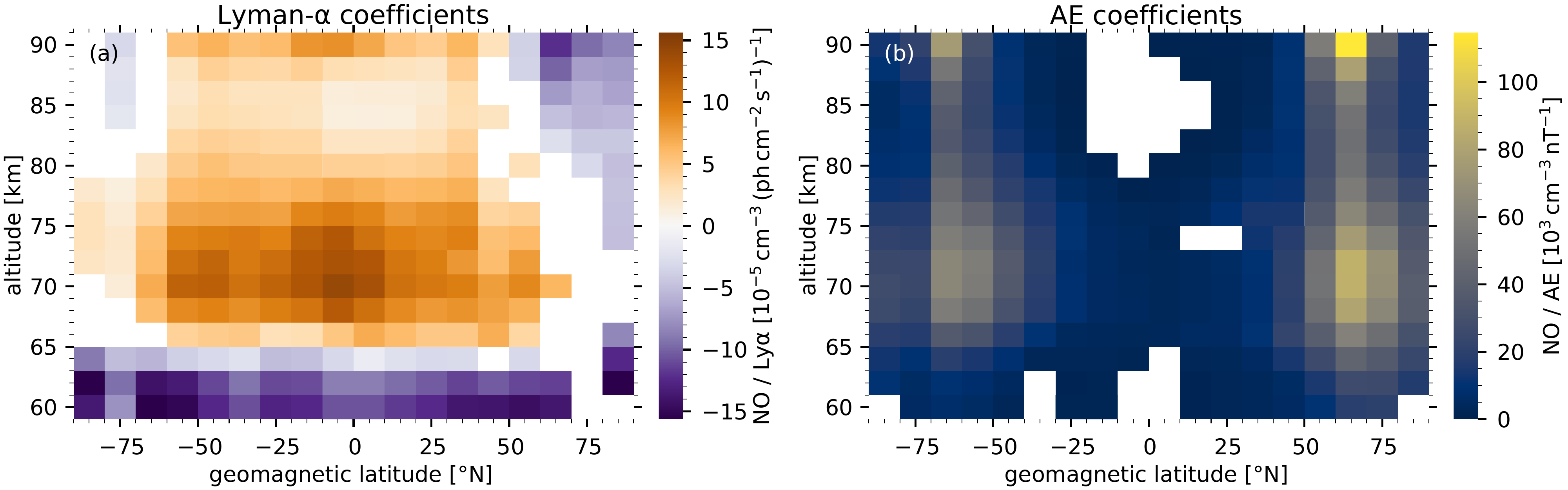}
\caption{Latitude--altitude distributions of the fitted
solar index parameter (Lyman-$\alpha$, \textbf{a}) and the
geomagnetic index parameter (AE, \textbf{b}) from
the non-linear model.
}\label{fig:nonlin.params}
\end{figure*}

Using the non-linear model, we show the latitude--altitude
distributions of the medians of the sampled
Lyman-$\alpha$ and geomagnetic
index coefficients in Fig.~\ref{fig:nonlin.params}.
The white regions indicate
values outside of the 95\% confidence region
or whose sampled distribution has a skewness larger than 0.33.
The MCMC method samples the parameter probability
distributions. Since we require the geomagnetic index
and constant lifetime parameters to be larger than zero (see Table~\ref{tab:paramspace}),
these sampled distributions are sometimes skewed towards zero
even though
the 95\% credible region is still larger than zero.
Excluding heavily skewed distributions avoids those cases
because the
``true'' parameter is apparently zero.

The Lyman-$\alpha$ parameter distribution shows that its
largest influence is at middle and low latitudes between
65 and 80\,km.
Another increase of the Lyman-$\alpha$
coefficient is indicated at higher altitudes above 90\,km.
The penetration of Lyman-$\alpha$ radiation decreases with
decreasing altitude as a result of scattering and absorption
by air molecules.
On the other hand, the concentration of air decreases with altitude.
At this stage we do not have an unambiguous explanation of this behaviour,
but it may be related to reaction pathways as laid out by~\citet{Pendleton1983},
which would relate the \chem{NO} concentrations to the \chem{CO_2}
and \chem{H_2O} (or \chem{OH}, respectively) profiles.
The Lyman-$\alpha$ coefficients are all negative below 65\,km.
We also observe negative values at high northern
latitude at all altitudes and at high southern latitudes
above 85\,km.
These negative coefficients indicate that
\chem{NO} photodissociation or conversion to other
species outweighs its production via UV radiation in those places.
The north--south asymmetry may be related to sampling
and the difference in illumination with respect to
geomagnetic latitudes; see Sect.~\ref{ssec:res.fit}.

The geomagnetic influence is largest at high latitudes
between 50 and 75\degree\ above about 65\,km.
The AE coefficients peak at around 72\,km and indicate
a further increase above 90\,km.
This pattern of the geomagnetic influence matches
the one found in~\citet{Sinnhuber2016}.
Unfortunately both increased influences above 90\,km
in Lyman-$\alpha$ and AE
cannot be studied at higher latitudes due to a large a priori contribution
to the data.

\begin{figure*}[t]
\centering
\includegraphics[width=16.6cm]{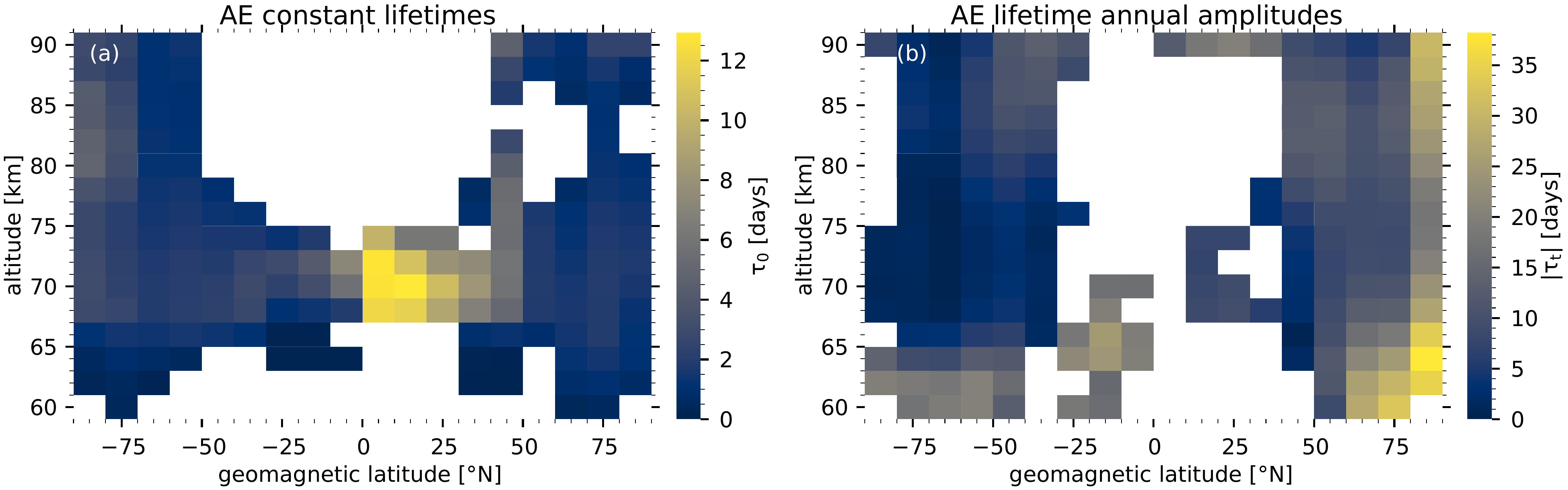}
\caption{Latitude--altitude distributions of the fitted
	base lifetime $\tau_0$ \textbf{(a)} and the amplitude of the
	annual variation $\lvert\tau_t\rvert$ \textbf{(b)}
	from the non-linear model.
}\label{fig:nonlin.tauparams}
\end{figure*}

The latitude--altitude distributions of the lifetime parameters
are shown in Fig.~\ref{fig:nonlin.tauparams}.
All values shown are within the
95\% confidence region.
As for the coefficients above, we also exclude regions where
the skewness was larger than 0.33.

The constant part of the lifetime, $\tau_0$, is below
2 days in most bins, except for exceptionally large
values ($> 10$~days) at low latitudes (0--20\degree N) between 68 and 74\,km.
Although we constrained the lifetime with an exponential
prior distribution, these large values apparently resulted
in a better fit to the data.
One explanation could be that because of the small geomagnetic
influence (the AE coefficient is small in this region),
the lifetime is more or less irrelevant.
The amplitude of the annual variation
($\lvert\tau_{\text{t}}\rvert
= \sqrt{\tau_{\text{cos}}^2 + \tau_{\text{sin}}^2}
= \sqrt{d^2 + e^2}$; see Eq.~\eqref{eq:mod.tau.t})
is largest at high latitudes in the Northern Hemisphere
and at middle latitudes in the Southern Hemisphere.
This difference could be linked to the geomagnetic latitudes
which include a wider range of geographic latitudes in the Southern
Hemisphere compared to the Northern Hemisphere.
Therefore, the annual variation is less apparent in the Southern Hemisphere.
The amplitude also increases with decreasing altitude below 75\,km
at middle and high latitudes
and with increasing altitude above that.
The increasing annual variation at low altitudes
can be the result of transport processes that
are not explicitly treated in our approach.
Note that the term \emph{lifetime} is not
a pure (photo)chemical lifetime; rather it indicates
how long the AE signal persists in the \chem{NO} densities.
In that sense it combines the
(photo)chemical lifetime with transport effects
as discussed in~\citet{Sinnhuber2016}.

\subsection{Parameter profiles}\label{ssec:res.prof}

\begin{figure*}[t]
\centering
\includegraphics[width=13cm]{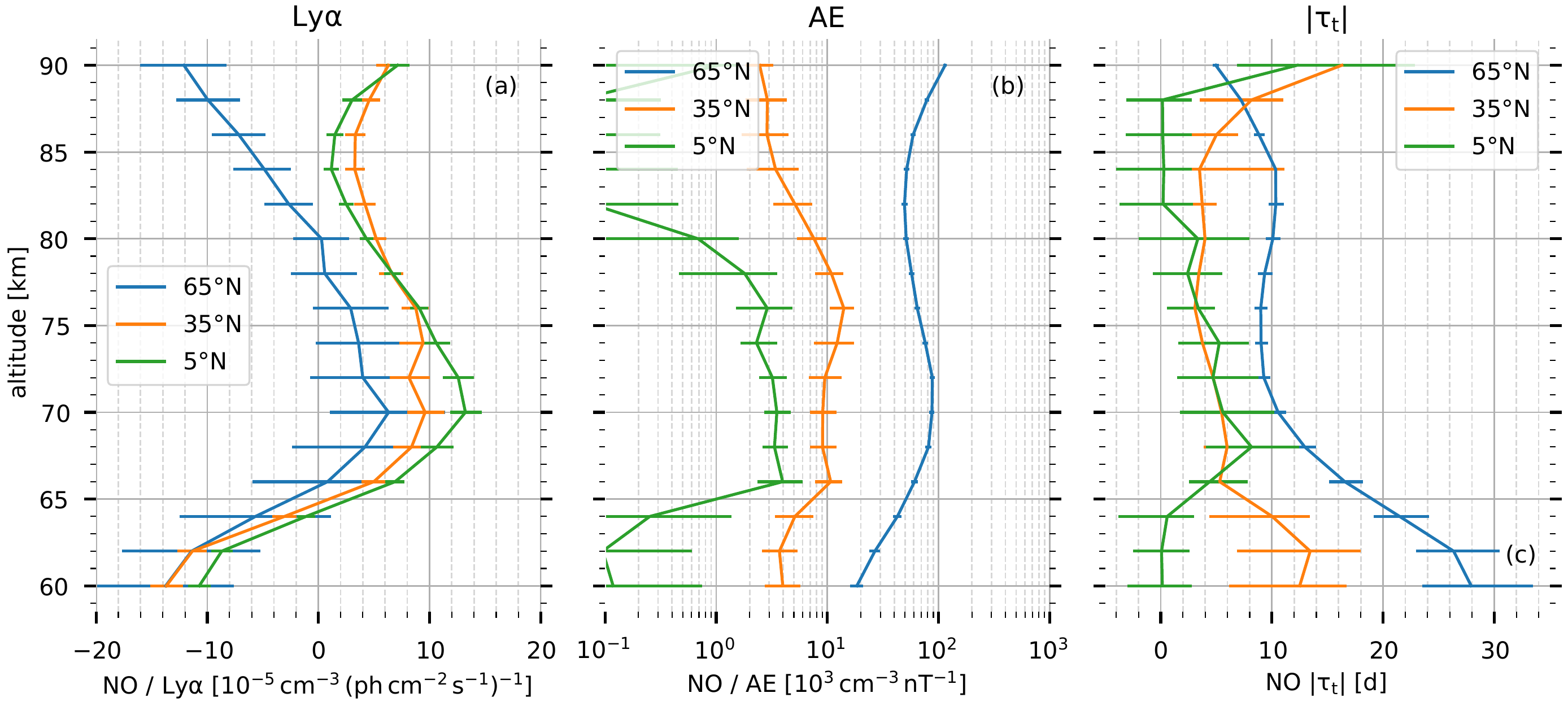}
\caption{Coefficient profiles of the
solar index parameter (Lyman-$\alpha$, left, \textbf{a}),
the geomagnetic index parameter (AE, middle, \textbf{b}),
and the amplitude of the annual variation of the \chem{NO} lifetime (right, \textbf{c})
at 5\degree N (green), 35\degree N (orange), and 65\degree N (blue).
The solid line indicates the median, and
the error bars indicate the 95\% confidence region.}\label{fig:prof.NH}
\end{figure*}

For three selected latitude bins in the Northern Hemisphere
(5, 35, and 65\degree N)
we present profiles of the fitted parameters in Fig.~\ref{fig:prof.NH}.
The solid line indicates the median, and
the error bars indicate the 95\% confidence region.
As indicated in Fig.~\ref{fig:nonlin.params}, the solar radiation
influence is largest between 65 and 80\,km.
Its influence is also up to a factor of 2
larger at low and middle latitudes compared to high latitudes,
where the coefficient only differs significantly from zero
below 65 and above 82\,km.
Similarly, the geomagnetic impact decreases with decreasing latitude by
1 order of magnitude from high to middle latitudes and at least
a further factor of 5 to lower latitudes.
The largest impact is around 70--72\,km and possibly above 90\,km
at high latitudes and is approximately constant between 66 and
76\,km at middle and low latitudes.
Note that the scale in Fig.~\ref{fig:prof.NH}b
is logarithmic.
The lifetime variation shows that at high latitudes,
geomagnetically affected \chem{NO} persists longer during winter
(the phase is close to zero for all altitudes at 65\degree N,
not shown here).
It persists up to 10 days longer between 85 and 70\,km
and increasingly longer below, reaching 28 days at 60\,km.

\begin{figure*}[t]
\centering
\includegraphics[width=13cm]{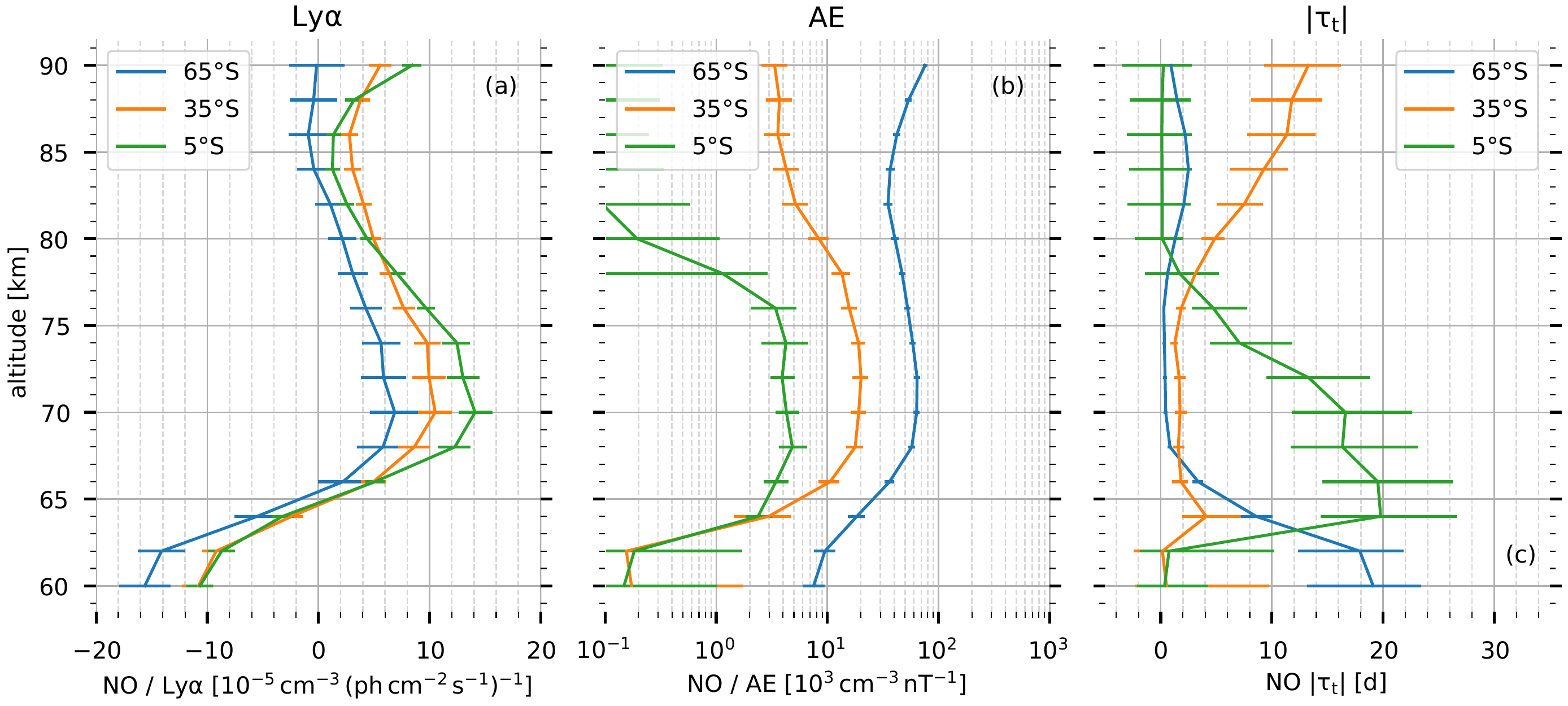}
\caption{Coefficient profiles of the
solar index parameter (Lyman-$\alpha$, left, \textbf{a}),
the geomagnetic index parameter (AE, middle, \textbf{b}),
and the amplitude of the annual variation of the \chem{NO} lifetime (right, \textbf{c})
at 5\degree S (green), 35\degree S (orange), and 65\degree S (blue).
The solid line indicates the median, and
the error bars indicate the 95\% confidence region.}\label{fig:prof.SH}
\end{figure*}

For the same latitude bins in the Southern Hemisphere
(5\degree S, 35\degree S, and 65\degree S)
we present profiles of the fitted parameters in Fig.~\ref{fig:prof.SH}.
Similar to the coefficients in the Northern Hemisphere (see Fig.~\ref{fig:prof.NH}),
the solar radiation influence is largest between 65\,km and 80\,km and also up
to a factor of two larger at low and middle latitudes compared to high latitudes.
However, the Lyman-$\alpha$ coefficients at 65\degree S are significant below 82\,km.
Also the geomagnetic AE coefficients show a similar pattern in
the Southern Hemisphere compared to the Northern Hemisphere,
decreasing by orders of magnitude from high to low latitudes.
Note that the AE coefficients at high latitudes are slightly
lower than in the Northern Hemisphere, whereas the coefficients
at middle and low latitudes are slightly larger.
This slight asymmetry was also found in the study
by~\cite{Sinnhuber2016} and may be related to AE being derived
solely from stations in the Northern Hemisphere~\citep{Mandea2011}.
With respect to latitude,
the annual variation of the lifetime seems to be reversed compared
to the Northern Hemisphere, with almost no variation at high latitudes
and longer persisting \chem{NO} at low latitudes.
A faster descent in the southern polar vortex may be responsible
for the short lifetime at high southern latitudes.
Another reason may be the mixture of air from inside and outside of
the polar vertex when averaging along geomagnetic latitudes
since the 65\degree S geomagnetic latitude band includes
geographic locations from about 45\degree S to 85\degree S.
A third possibility may be the exclusion of the Southern Atlantic Anomaly
from the retrieval~\citep{Bender2013, Bender2017}
where presumably the particle-induced impact on \chem{NO} is largest.

\subsection{Discussion}\label{ssec:res.discuss}

The distribution of the parameters confirms our understanding
of the processes producing \chem{NO} in the mesosphere to a
large extent. The Lyman-$\alpha$ coefficients are related to
radiative processes such as production by
UV or soft X-rays, either directly or via intermediary
of photoelectrons. The photons are not influenced by Earth's
magnetic field, and the influence of these processes is largest
at low latitudes and decreases towards higher latitudes.
We observe negative Lyman-$\alpha$ coefficients
below 65\,km at all latitudes and at high
northern latitudes above 80\,km.
These negative Lyman-$\alpha$ coefficients
indicate that at high solar activity,
photodissociation by $\lambda < 191\,$nm photons,
photoionization by $\lambda < 134\,$nm photons,
or collisional loss and conversion to other species
outweigh the production from higher energy photons ($<40$\,nm).
At high southern latitudes these negative Lyman-$\alpha$
coefficients are not as pronounced as at high northern latitudes.
As mentioned in Sect.~\ref{ssec:res.morph}, this north--south
asymmetry may be related to sampling
and the difference in illumination with respect to
geomagnetic latitudes, see also Sect.~\ref{ssec:res.fit}.

The AE coefficients are largest at auroral latitudes as
expected for the particle nature of the associated \chem{NO} production.
The AE coefficient can be considered an effective production rate
modulated by all short-term ($\ll 1$\,day) processes.
To roughly estimate this production rate, we divided the
coefficient of the (daily) AE by 86400\,s which follows
the approach in~\citet{Sinnhuber2016}.
We find a maximum production rate of about
1\,cm$^{-3}$\,nT$^{-1}$\,s$^{-1}$ around 70--72\,km.
This production rate also agrees with the one estimated
by~\citet{Sinnhuber2016} by a superposed epoch analysis
of summertime \chem{NO}.
Comparing the \chem{NO} production to the ionization rates
from~\citet{Verronen2013} from 1 to 3~January~2005
(assuming approximately 1 \chem{NO} molecule per ion pair), our
model overestimates the ionization derived from AE on these days.
The AE values of 105, 355, and 435\,nT translate
to 105, 355, and 435\,\chem{NO} molecules\,cm$^{-3}$\,s$^{-1}$, about 4 times
larger than would be estimated from the ionization rates in~\citet{Verronen2013} but
agreeing with~\citet{Sinnhuber2016}.
The factor of 4 may be related to the slightly different locations,
around 60\degree N~\citep{Verronen2013} compared to around 65\degree N
here and in~\citet{Sinnhuber2016}, in which the ionization rates may be higher.

The associated constant part $\tau_0$ of the lifetime ranges
from around 1 to around 4 days, except for large $\tau_0$ at
low latitudes around 70\,km.
As already discussed in Sect.~\ref{ssec:res.morph},
these large lifetimes may be a
side effect of the small geomagnetic coefficients and more
or less arbitrary.
The magnitude is similar to what was found in the
study by~\citet{Sinnhuber2016} using only the summer data.

The annual variation of the lifetime is largest at high
northern latitudes with a nearly constant amplitude of
10~days between 70 and 85\,km.
An empirical lifetime of 10~days in winter was used by~\citet{Sinnhuber2016}
to extend the \chem{NO} predicted by the summer analysis
to the larger values in winter.
Here we could confirm that 10~days is a good approximation
of the \chem{NO} lifetime in winter, but it varies with altitude.
The altitude distribution agrees with the increasing
photochemical lifetime at large solar zenith angles~\citep[Fig.~7b]{Sinnhuber2016}.
The larger values in our study are similarly related to
transport and mixing effects which alter the observed lifetime.
The small variation of the lifetime at high southern
latitudes could be a sampling issue because
SCIAMACHY only observes small variations there in winter
(see Figs.~\ref{fig:lin.fit} and~\ref{fig:nonlin.fit}).
Note that the results (in particular the large annual variation)
in the northernmost latitude bin should be taken with caution
because this bin is sparsely sampled by SCIAMACHY, and
the large winter \chem{NO} concentrations are
actually absent from the data.

\conclusions\label{sec:conclusions}  
We propose an empirical model to estimate the \chem{NO} density in the mesosphere
(60--90\,km) derived from measurements from SCIAMACHY nominal-mode limb scans.
Our model calculates \chem{NO} number densities for geomagnetic
latitudes using the solar Lyman-$\alpha$ index and the geomagnetic
AE index.
Two approaches were tested, a linear approach containing annual and
semi-annual harmonics and a non-linear version using a finite
and variable lifetime for the geomagnetically induced variations.
From our proposed models,
the linear variant only describes part of the \chem{NO} variations.
It can describe the summer variations but
underestimates the large number densities during winter.
The non-linear version derived from the SCIAMACHY \chem{NO} data
describes both variations using an annually varying finite lifetime
for the particle-induced \chem{NO}.
However, in cases of dynamic disturbances of the mesosphere,
for example in early 2004 or in early 2006,
the indirectly enhanced
\chem{NO}~\citep[see, for example,][]{Randall2007} is not
captured by the model.
These remaining variations are treated as statistical noise.

\citet{Sinnhuber2016} use a superposed epoch analysis limited to
the polar summer to model the \chem{NO} data.
Here we extend that analysis to use all available SCIAMACHY
nominal-mode \chem{NO} data for all seasons.
However, during summer the present results show comparable
\chem{NO} production per AE unit and similar lifetimes to
the~\citet{Sinnhuber2016} study.

The parameter distributions indicate in which regions the different processes
are significant.
We find that these distributions match our current understanding of
the processes producing and depleting \chem{NO} in the
mesosphere~\citep{Funke2014a, Funke2014, Funke2016, Sinnhuber2016, Hendrickx2017, Kiviranta2018}.
In particular, the influence of Lyman-$\alpha$
(or solar UV radiation in general) is largest at low and middle latitudes,
which is explained by the direct production of \chem{NO} via solar UV
or soft X-ray radiation~\citep{Barth1988, Barth2003}.
The geomagnetic influence is largest at high latitudes and is best explained by the
production from charged particles that enter the atmosphere
in the polar regions along the magnetic field.

A potential improvement would be to use actual measurements of precipitating particles
instead of the AE index.
Using measured fluxes could help to confirm our current understanding
of how those fluxes relate to ionization~\citep{Turunen2009, Verronen2013}
and subsequent \chem{NO} production~\citep{Sinnhuber2016}.
Furthermore, including dynamical transport, as for example
in~\citet{Funke2016}, could improve our knowledge of the
combined direct and indirect \chem{NO} production in the mesosphere.

\authorcontribution{SB developed the model, prepared and performed the
data analysis and set up the manuscript, MS provided input on the model
and the idea of a variable \chem{NO} lifetime.
JPB and PJE contributed to the discussion and use of language.
All authors contributed to
the interpretation and discussion of the method and the results as well as
to the writing of the paper.}

\codedataavailability{The SCIAMACHY \chem{NO} data set used in this study
	is available at \path{https://zenodo.org/record/1009078}
	\citep{Bender2017c}. The python code to prepare the data (daily zonal
	averaging) and to perform the analysis is available at
	\path{https://zenodo.org/record/1401370} \citep{Bender2018a} or at
	\path{https://github.com/st-bender/sciapy}. The daily zonal
	mean \chem{NO} data and the sampled parameter distributions
	are available at
	\path{https://zenodo.org/record/1342701} \citep{Bender2018b}.
	The solar Lyman-$\alpha$ index data were downloaded from
	\path{http://lasp.colorado.edu/lisird/data/composite_lyman_alpha/},
	the AE index data were downloaded from \path{http://wdc.kugi.kyoto-u.ac.jp/aedir/},
	and the daily mean values used in this study are available within the
	aforementioned data set.
}

\begin{acknowledgements}
S.B.\ and M.S.\ thank the Helmholtz-society for funding part of this project
under the grant number VH-NG-624.
S.B.\ and P.J.E.\ acknowledge support from the Birkeland Center
for Space Sciences (BCSS), supported by the Research Council of
Norway under the grant number 223252/F50.
The SCIAMACHY project was a national contribution to the ESA Envisat,
funded by German Aerospace (DLR),
the Dutch Space Agency, SNO, and the Belgium ministry.
The University of Bremen as Principal Investigator has led the scientific
support and development of the SCIAMACHY instrument as well as the scientific exploitation of its
data products.
This work was performed on the Abel Cluster,
owned by the University of Oslo and Uninett/Sigma2,
and operated by the Department for Research Computing at USIT,
the University of Oslo IT-department. http://www.hpc.uio.no/
\end{acknowledgements}


\bibliography{bigbib}{}
\bibliographystyle{copernicus}

\end{document}